# REGRESSION WITH MISSING Y'S:

# AN IMPROVED STRATEGY

# FOR ANALYZING MULTIPLY IMPUTED DATA

Paul T. von Hippel

*Correspondence*: Paul T. von Hippel, Department of Sociology, Ohio State University, 300 Bricker Hall, 190 N. Oval Mall, Columbus OH 43210, <u>von-hippel.1@osu.edu</u>, 614 688-3768.

I thank Paul Allison, Doug Downey, Jerry Reiter, and Donald Rubin for helpful feedback on an earlier draft.

**Abstract**

When fitting a generalized linear model—such as a linear regression, a logistic regression, or a hierarchical linear model—analysts often wonder how to handle missing values of the dependent variable $Y$. If missing values have been filled in using multiple imputation, the usual advice is to use the imputed $Y$ values in analysis. We show, however, that using imputed $Y$s can add needless noise to the estimates. Better estimates can usually be obtained using a modified strategy that we call *multiple imputation, then deletion* (MID). Under MID, all cases are used for imputation, but following imputation cases with imputed $Y$ values are excluded from the analysis. When there is something wrong with the imputed $Y$ values, MID protects the estimates from the problematic imputations. And when the imputed $Y$ values are acceptable, MID usually offers somewhat more efficient estimates than an ordinary MI strategy.

# 1. MULTIPLE IMPUTATION, THEN DELETION (MID)

*Multiple imputation* (MI) is an increasingly popular tool for analyzing data with missing values (Rubin 1987). As it is commonly used, MI is part of a four-step estimation strategy:

1. *Replication*. Make multiple copies of the incomplete data set.

2. *Imputation*. In each copy, replace each missing value with a plausible random imputation. (Imputations are drawn conditionally on the observed values of all of the variables.)

3. *Analysis*. Analyze each imputed data set separately, using the standard methods that are used for complete data.

4. *Recombination*. Combine the results of the separate analyses, using formulas that account for variation within and between the imputed data sets.

Researchers often use MI when they are estimating the conditional distribution of an outcome $Y$ given some inputs $X=(X_1,...,X_p)$. For example, analysts may use MI in estimating the parameters of a generalized linear model such as a normal or logistic regression, or a hierarchical linear model.

Researchers often ask how they should handle the dependent variable $Y$. The easy question is whether $Y$ should be used to impute $X$. The answer is yes (e.g., Allison 2002). If $Y$ is not used to impute $X$, then $X$ will be imputed as though it has no relationship to $Y$. When the imputed data are analyzed, the estimated slope of $Y$ on $X$ will be biased toward zero, since a value of zero was tacitly assumed in imputation.[1]

This paper focuses on a harder question, which is what to do with cases that are missing $Y$. The answer begins with two remarks in Little's (1992) paper "Regression with Missing $X$'s," p. 1227:

1.   "If the $X$'s are complete and the missing values of $Y$ are missing at random, then the incomplete cases contribute no information to the regression of $Y$ on $X_1,...,X_p$."

In other words, when the $X$'s are complete, there is no need for imputation, because maximum-likelihood estimates can be obtained simply by deleting the cases with missing $Y$. Using imputed $Y$ values in analysis would simply add noise to these estimates. On the other hand, Little continues, p. 1227,

2.   "if values of $X$ are missing as well as $Y$, then cases with $Y$ missing can provide a minor amount of information for the regression of interest, by improving prediction of missing $X$'s for cases with $Y$ present."

_____________________

[1] Imputers sometimes worry that, by including $Y$ in the imputation step, they are assuming something unwarranted about the $X$-$Y$ relationship. This concern is misplaced. By including $Y$ in the imputations, you are not assuming that $Y$ has any particular relationship with $X$; the relationship could be positive, negative or zero, and any of these possibilities will be accounted for by the imputation model. On the other hand, if you exclude $Y$ from the imputations, you *are* making an assumption. You are assuming that there is no direct relationship between $X$ and $Y$.

This second remark implies that cases with missing *Y* should be used in the imputation step, since those cases may contain information that is useful for imputing *X* in other cases. But after imputation, cases with imputed *Y* have nothing more to contribute; when the data are analyzed, random variation in the imputed *Y* values adds nothing but noise to the estimates.

In short, cases with missing *Y* are useful for imputation, but not for analysis.

In light of this observation, we propose a new estimation strategy that we call *multiple imputation, then deletion* (MID). MID is just like a conventional MI strategy except that cases with imputed *Y* are deleted before analysis:

1. *Replication.*

2. *Imputation.*

    **2½. *Deletion*. Delete all cases that have imputed values for *Y*.**

3. *Analysis.*

4. *Recombination.*

One advantage of MID is efficiency. Compared to an ordinary MI strategy (one that retains imputed *Y*s), MID tends to give less variable point estimates, more accurate standard-error estimates, and shorter confidence intervals with equal or higher coverage rates. To state these advantages in terms of hypothesis tests, MID tests tend to have greater power while maintaining equal or lower significance levels. MID's advantage in efficiency is often minor, but it can be substantial when there are a lot of missing values and relatively few imputed data sets.

A second and perhaps more important advantage is that MID is robust to problems in the imputation model. Problems in imputing $Y$ cannot affect MID estimates, because cases with imputed $Y$ are deleted before analysis. Problems in imputing $X$ may also have little effect if, as is often the case, missing $X$s tend to occur in the same cases as missing $Y$s.

The importance of deleting problematic imputations bears some emphasis because, in practice, there are several things that can go wrong when data are imputed. For example, nonlinear relationships, such as interactions, may be carelessly omitted from the imputation model (Allison 2002). And if the imputation model is specified carefully, the imputation software may have undocumented biases (Allison 2000; von Hippel 2004). Even if the software works well, it may have limited flexibility, so that the analyst has to impute skewed or discrete variables as though they were normal. An inappropriate assumption of normality can result in unrealistic imputations such as a negative body weight, or a dummy variable with a value of 0.6. To "fix" such unrealistic values, some analysts round or transform imputed values—but these fixes can introduce biases of their own (von Hippel under review; Horton, Lipsitz, and Parzen 2003; Allison 2005).

In short, since imputation can go wrong in several ways, it seems desirable to reduce reliance on imputed values. MID does this.

Like any missing-data method, MID relies on certain assumptions. In particular, MID assumes that missing $Y$ values are *ignorable* (Little & Rubin 2002) in the sense that the unobserved $Y$ values are similar to observed $Y$ values from cases with similar values for $X$. In addition, MID assumes that missing $X$ values are ignorable in cases with missing $Y$. The assumption of ignorability is not unique to MID; in fact, the vast majority of conventional MI analyses assume that missing values are ignorable. But there are extensions of MI that relax the assumption of ignorability (Rubin 1987). Though these extensions are rarely used and not always effective (Rubin 2003), under MID they can hardly be used at all.[2]

MID also relies on the assumption that the imputed $Y$ values contain no useful information. This assumption is usually valid, but in some data sets the imputed $Y$ values have been enriched by auxiliary information from an outside source. As we will show in section 7, hosever, the "signal" in such auxiliary information must be quite strong before it overcomes the "noise" from the random component in the imputed $Y$ values.

In short, although there are special circumstances when a conventional MI strategy is superior to MID, under most practical circumstances MID has at least a small advantage.

In this paper, we describe applications of MID in social research (section 2); we explain why MID works (section 3); and we demonstrate the efficiency of MID estimates both analytically (section 4) and through simulations (section 5). In the final sections of the paper, we outline some extensions and limitations of MID (sections 6-7), and argue that the limitations are usually minor compared to the advantages.

---

[2] Imputation techniques for nonignorable missingness can be used with MID, but only under the contrived assumption that the nonignorability is confined to cases where $X$ is missing and $Y$ is not.

## 2. EXAMPLES OF MID IN SOCIAL RESEARCH

Two recent pieces of social research highlight the advantages of MID. In these examples, MID was especially helpful because many values were missing and it was difficult to specify a convincing imputation model for $Y$.

Chamberlain et al. (2005) used MID in a model of sexual harassment $Y$ at 204 firms. Nearly half of all $Y$ values were missing, and imputation was tricky because $Y$ was a categorical variable, with four possible values. Although there does exist software for imputing multi-category variables (e.g., Raghunathan et al. 2002; Royston 2004), initially it was more convenient to use imputation software that assumed normality (the MI procedure in SAS version 9). In preparing the data, the researchers recoded the categorical $Y$ variable as a set of dummy variables (0/1), and imputed each dummy as though it were normal. The normal imputation model filled in the dummy variables with nonsense—values other than 0 and 1—but since the imputed $Y$s were then deleted, their unrealistic values had no effect on the analysis. Later, the analyses were repeated using categorical imputation software, and the results were not materially different.

Downey et al. (2004) used MID in their longitudinal analysis of data from the Early Childhood Longitudinal Study, Kindergarten cohort, or ECLS-K (National Center for Education Statistics 2002). The ECLS-K is a cluster sample with approximately 20,000 children clustered in 1,000 schools. Unfortunately, available imputation software offers no convenient way to account for clustering,[3] so Downey et al. (2004) used an imputation model that ignored the clusters. Again, though, since many of the imputed values were deleted before analysis, the misspecification of the imputation model had little effect on the results.

## 3. WHY DOES MID WORK?

MID works because, after imputation, cases with imputed $Y$ contain no information about the regression of interest (cf. Little 1992). A conventional MI strategy tries to estimate the information in cases with imputed $Y$, but the attempt yields only estimation error. An MID strategy, by contrast, simply discards cases with imputed $Y$, recognizing that there is no information to be found in them.

### 3.1   Why cases with imputed $Y$ contain no information

Cases with imputed $Y$ quite literally contain no information about the regression of $Y$ on $X$. In statistics, *information* is a synonym for the log of the likelihood (Kullback 1959),[4] and the log-likelihood for cases with imputed $Y$ is exactly zero.

---

[3] Rubin (1996) suggests using an indicator variable for each cluster; however, most imputation software would treat the cluster indicators as fixed effects, which leads to estimation problems in the ECLS-K since the clusters are small with many missing values (Reiter and Raghunathan). Schafer and Yucel (2002) develop a method that accounts for clustering in a random-effects framework; however, their method is implemented in an outdated version of S-Plus, and assumes that only $Y$ (not $X$) has missing values.

To see this, let $(X,Y) = (x_i, y_i)$, $i=1,\ldots,n$, be a data matrix representing $n$ observations on a univariate outcome $y_i$ and a vector of inputs $x_i = (x_{i1}, x_{i2}, \ldots)$, and suppose we wish to estimate the parameter vector $\theta_{Y/X}$, or more simply $\theta$, that governs the conditional distribution of $Y$ given $X$. In a regression model, for example, the parameter vector would be $\theta = (\alpha\ \beta_1\ \beta_2\ \ldots\ \sigma^2)^T$ where $\alpha$ is the intercept, $\beta_1, \beta_2,\ldots$ are the slopes, and $\sigma^2$ is the residual variance.

If there were no missing values, then the likelihood of $\theta$ would be

$$L(\theta \mid X, Y) = f(Y \mid X, \theta) \tag{1},$$

where $f$ is the conditional distribution of $Y$ given $X$ and $\theta$.

When some values are missing, we need a little more notation. Following Rubin (1987), let $Y_{obs}$ and $Y_{mis}$ be the observed and missing values of $Y$, and let $X_{obs}$ and $X_{mis}$ be the observed and missing values of $X$. Then the likelihood of $\theta$ given the observed $X$ and $Y$ values can be obtained by integrating over all possible values of the missing data,

$$L(\theta \mid X_{obs}, Y_{obs}) = \int L(\theta \mid X_{obs}, Y_{obs}, X_{mis}, Y_{mis}) p(X_{mis}, Y_{mis} \mid X_{obs}, Y_{obs}) dX_{mis} dY_{mis} \tag{2}.$$

Here each possible value of the missing data $(X_{mis}, Y_{mis})$ is weighted according to the posterior predictive distribution $p(X_{mis}, Y_{mis} \mid X_{obs}, Y_{obs})$ of the missing values given the observed values. The assumption that the posterior predictive distribution depends only on

---

[4] *Information* also refers to a way of approximating the log likelihood by taking its second derivative near the maximum likelihood estimate.

the observed values is known as _**ignorability**_. (We will discuss ignorability further in section 7.)

To understand why the cases with missing $Y$ contribute nothing to this likelihood, we break the data set into cases with $Y$ missing and cases with $Y$ observed. In the cases with $Y$ missing, let $X_{obs/Ymis}$ and $X_{mis/Ymis}$ be the observed and missing values of $X$; likewise, in cases with $Y$ observed, let the observed and missing $X$ values be $X_{obs/Yobs}$ and $X_{mis/Yobs}$. Then the total likelihood $L(\boldsymbol{\theta} \mid X_{obs}, Y_{obs})$ can be broken into two components: (1) the likelihood for cases with observed $Y$ values and (2) the likelihood for cases where $Y$ is missing:

$$L(\boldsymbol{\theta} \mid X_{obs}, Y_{obs}) = L_{Yobs}(\boldsymbol{\theta} \mid X_{obs/Yobs}, Y_{obs}) L_{Ymis}(\boldsymbol{\theta} \mid X_{obs/Ymis}) \qquad (3).$$

Now the likelihood for cases with $Y$ missing must equal one, since the likelihood in these cases is the conditional $Y$ distribution integrated over all possible $Y$ values:

$$L_{Ymis}(\boldsymbol{\theta} \mid X_{obs/Ymis}) = \int f(Y_{mis} \mid X_{obs/Ymis}, \boldsymbol{\theta}) dY_{mis} = 1 \qquad (4).$$

Since the log of one is zero, it follows that the log-likelihood of these cases is zero:

$$\log L_{Ymis}(\boldsymbol{\theta} \mid X_{obs/Ymis}) = 0 \qquad (5).$$

That is, the cases with missing $Y$ contain no *information* about the regression parameters, as we claimed earlier.

It follows that the overall likelihood is simply the likelihood for cases with $Y$ observed:

$$L(\boldsymbol{\theta} \mid X_{obs}, Y_{obs}) = L_{Yobs}(\boldsymbol{\theta} \mid X_{obs/Yobs}, Y_{obs}) \qquad (6).$$

This does not mean that cases with missing $Y$ are useless. As Little (1992, p. 1227) writes, "cases with $Y$ missing can provide a minor amount of information for the regression of interest, by improving prediction of missing $X$'s for cases with $Y$ present." More precisely, the likelihood for cases with $Y$ present is

$$L_{Yobs}(\boldsymbol{\theta} \mid X_{obs}, Y_{obs}) = \int f(Y_{obs} \mid X_{obs|Yobs}, X_{mis|Yobs}) p(X_{mis|Yobs} \mid Y_{obs}, X_{obs|Yobs}) dX_{mis|Yobs} \quad (7),$$

and the cases with $Y$ missing play a role in the posterior predictive distribution $p(X_{mis|Yobs} \mid Y_{obs}, X_{obs|Yobs})$ of the $X$ values that are missing from cases with $Y$ observed.

## 3.2 Estimating the likelihood under MI

Multiple imputation can be justified as a way of estimating the likelihood by approximating the integral in (2) (Little and Rubin 2002). Instead of integrating across all possible missing values, we average across a few possible values. More specifically, we make $M$ copies of the incomplete data set, and in the $m^{\text{th}}$ copy ($m$=1,2,…) we fill in the missing $X$ and $Y$ values with a set of imputations ($X_{mis}^{(m)}$, $Y_{mis}^{(m)}$) drawn from an estimate of the posterior predictive distribution $p(X_{mis}, Y_{mis} \mid X_{obs}, Y_{obs})$. In the $m^{\text{th}}$ imputed data set, we estimate the likelihood $\hat{L}^{(m)}(\boldsymbol{\theta} \mid X_{mis}^{(m)}, X_{obs}, Y_{mis}^{(m)}, Y_{obs})$ of $\boldsymbol{\theta}$ given the imputed values:

$$\hat{L}^{(m)}(\boldsymbol{\theta} \mid X_{mis}^{(m)}, X_{obs}, Y_{mis}^{(m)}, Y_{obs}) = f(Y_{mis}^{(m)}, Y_{obs} \mid X_{mis}^{(m)}, X_{obs}, \boldsymbol{\theta}) \quad (8).$$

Finally, we average the estimated likelihoods across the $M$ imputed data sets:

$$\hat{L}_{M,MI}(\boldsymbol{\theta} \mid X_{obs}, Y_{obs}) = \frac{1}{M}\sum_{m=1}^{M}\hat{L}^{(m)}(\boldsymbol{\theta} \mid X_{mis}^{(m)}, X_{obs}, Y_{mis}^{(m)}, Y_{obs}) \tag{9}.$$

This is the MI estimate for the likelihood of $\boldsymbol{\theta}$.

## 3.3   Estimating the likelihood under MID

MID estimates the same likelihood, but it does so more efficiently. Since the overall likelihood is just the likelihood for cases with $Y$ observed,

$$L(\boldsymbol{\theta} \mid X_{obs}, Y_{obs}) = L_{Yobs}(\boldsymbol{\theta} \mid X_{obs\mid Yobs}, Y_{obs}) \tag{10},$$

it follows that the likelihood can be estimated without using imputed $Y$ values. In the $m^{\text{th}}$ imputed data set, the estimate of the likelihood using only the observed $Y$ values is

$$\hat{L}_{Yobs}^{(m)}(\boldsymbol{\theta} \mid X_{mis}^{(m)}, X_{obs}, Y_{mis}^{(m)}, Y_{obs}) = f(Y_{obs} \mid X_{mis\mid Yobs}^{(m)}, X_{obs}, \boldsymbol{\theta}) \tag{11}.$$

And averaging the estimated likelihoods across the $M$ imputed data sets gives

$$\hat{L}_{M,MID}(\boldsymbol{\theta} \mid X_{obs}, Y_{obs}) = \sum_{m=1}^{M}\hat{L}_{Yobs}^{(m)}(\boldsymbol{\theta} \mid X_{mis}^{(m)}, X_{obs}, Y_{obs}) \tag{12},$$

which is the MID estimate of the likelihood.

It follows that the estimated likelihood will usually be more accurate under MID than under MI. MI has an extra source of estimation error, because MI must estimate the likelihood for cases with missing $Y$. MID, by contrast, avoids this estimation error by simply assuming, correctly, that the exactly likelihood for cases with missing $Y$ is $L_{Ymis}(\boldsymbol{\theta} \mid X_{obs\mid Ymis}) = 1$.

# 4. ESTIMATION UNDER MID: MORE EFFICIENT POINT ESTIMATES, STANDARD ERRORS, AND CONFIDENCE INTERVALS

Some analysts are skeptical of the claim that MID increases the precision of statistical estimates. How can a procedure increase precision by throwing cases out? Doesn't deleting cases always *increase* the standard error?

It is true that deleting imputed *Y*s will increases the standard error *within* each imputed data set. But in compensation, by reducing the influence of randomly imputed values, MID reduces the amount that estimates vary *from one imputed data set to another*. As it turns out, it is better have a small variance *between* imputed data sets than it is to have a small variance *within* imputed data sets. The variance within imputed data sets can be estimated more accurately than the variance between imputed data sets, because the number of imputed data sets is typically much smaller than the number of cases within each.

Below we elaborate this argument by comparing the precision of point estimates, standard-error estimates, and confidence intervals under MI and MID. Not only does the discussion clarify the advantages of MID, it also presents explicit formulas for MID estimates, and clarifies the fine points of MID estimation.

## 4.1 Review of MI estimation

Suppose we want to estimate $\theta$, a scalar component of the parameter vector $\boldsymbol{\theta}$. For example, suppose that $\theta$ is the intercept or slope in a regression analysis. Under MI, $\theta$ is estimated as follows (Rubin 1987).

Analyze the $m^{\text{th}}$ imputed data set as though it were complete, and let $\hat{\theta}_{MI}^{(m)}$ and $\sqrt{\hat{W}_{MI}^{(m)}}$ be the *completed-data* estimates of the parameter and its standard error (Rubin 1987). In practice, these are usually maximum likelihood estimates, although for theoretical reasons it is helpful to think of them as the mean and standard deviation of the posterior density of $\theta$. (See section 4.3.1.)

An MI point estimate $\hat{\theta}_{M,MI}$ is obtained by averaging the completed-data point estimates across the $M$ imputed data sets:

$$\hat{\theta}_{M,MI} = \frac{1}{M}\sum_{m=1}^{M}\hat{\theta}_{MI}^{(m)} \tag{13}.$$

And an MI standard-error estimate $\sqrt{\hat{T}_{M,MI}}$ is obtained by adding the variances of the point estimate within and between the imputed data sets:

$$\sqrt{\hat{T}_{M,MI}} = \sqrt{\hat{W}_{M,MI} + \hat{B}_{M,MI}} \tag{14},$$

Here the within-imputation variance $\hat{W}_{M,MI}$ is the mean square of the completed-data standard errors across the $M$ imputed data sets,

$$\hat{W}_{M,MI} = \frac{1}{M}\sum_{m=1}^{M}\hat{W}_{MI}^{(m)} \tag{15},$$

and the between-imputation variance $\hat{B}_{M,MI}$ is the variance of the completed-data point estimate $\hat{\theta}_{MI}^{(m)}$ across the $M$ imputed data sets (Rubin 1987):[5]

$$\hat{B}_{M,MI} = \frac{1+1/M}{M-1} \sum_{m=1}^{M} (\hat{\theta}_{MI}^{(m)} - \hat{\theta}_{M,MI})^2 \qquad (16).$$

So $\hat{W}_{M,MI}$ and $\hat{B}_{M,MI}$ are the *within* and *between* variances, and their sum $\hat{T}_{M,MI}$ is the *total* variance.

The ratio of between to total variance is an estimate of the *fraction of missing information* $\hat{\gamma}_{M,MI}$:

$$\hat{\gamma}_{M,MI} = \frac{\hat{B}_{M,MI}}{\hat{T}_{M,MI}} \qquad (17).$$

A confidence interval for $\theta$ is

$$\hat{\theta}_{M,MI} \pm t_{\hat{v}_{M,MI}} \sqrt{\hat{T}_{M,MI}} \qquad (18),$$

where $t_{\hat{v}_{M,MI}}$ is a fractile from the $t$ distribution with $\hat{v}_{M,MI}$ degrees of freedom. The degrees of freedom $\hat{v}_{M,MI}$ is the harmonic total of two components (Barnard and Rubin 1999),

---

[5] Rubin's (1987) expression for $\hat{B}_{M,MI}$ omits the factor (1+1/$M$). To compensate, he includes this factor in his expression for the total variance, which he gives as $\hat{T}_{M,MI} = \hat{W}_{M,MI} + (1+1/M)\hat{B}_{M,MI}$.

$$\hat{v}_{M,MI} = \left( \frac{1}{\hat{v}_{imp,M,MI}} + \frac{1}{\hat{v}_{obs,MI}} \right)^{-1} \tag{19},$$

where the first component $v_{imp,M,MI}$ is the degrees of freedom in the imputations,

$$\hat{v}_{imp,M,MI} = (M-1)/\hat{\gamma}_{M,MI}^2 \tag{20},$$

and the second component is the degrees of freedom in the observed data,

$$\hat{v}_{obs,MI} = \frac{v_{com,MI}+1}{v_{com,MI}+3} v_{com,MI}(1-\hat{\gamma}_{M,MI}) \approx v_{com,MI}(1-\hat{\gamma}_{M,MI}) \tag{21}.$$

Here $v_{com,MI}$ is the degrees of freedom that would be observed if the data were complete. For example, in a bivariate linear regression with $n$ cases, $v_{com,MI} = n-2$.

## 4.2   MID estimation

MID estimation follow the same logic as ordinary MI estimation. MID simply uses fewer cases.

To use an MID strategy, impute the data as usual, then delete all the cases with imputed $Y$. Now obtain estimates using the MI formulas above, reducing the degrees of freedom to reflect the apparent reduction in sample size.

More formally, let $\hat{\theta}_{MID}^{(m)}$ and $\hat{W}_{MID}^{(m)}$ be the point estimate and squared standard error obtained by analyzing the $m^{\text{th}}$ imputed-then-deleted data set as though it were a complete data set with no missing values. Then MID estimates are obtained by substituting $\hat{\theta}_{MID}^{(m)}$ and $\hat{W}_{MID}^{(m)}$ for $\hat{\theta}_{MI}^{(m)}$ and $\hat{W}_{MI}^{(m)}$ into the MI formulas given above. These substitutions yield an MID point estimate $\hat{\theta}_{M,MID}$, an MID standard-error estimate $\sqrt{\hat{T}_{M,MID}}$, an MID fraction-of-missing information estimate $\hat{\gamma}_{M,MID}$, an MID degrees-of-freedom estimate $\hat{v}_{M,MID}$, and an MID confidence interval $\hat{\theta}_{M,MID} \pm t_{\hat{v}_{M,MID}} \sqrt{\hat{T}_{M,MID}}$.

All of the substitutions are straightforward. The only detail to spell out is what value to use for the complete-data degrees of freedom, represented by $v_{com,MI}$ under MI and by $v_{com,MID}$ under MID. Under MI, $v_{com,MI}$ is the degrees of freedom that would be observed if the entire data set were complete. Under MID, $v_{com,MID}$, is the degrees of freedom that would be observed if *only the cases with observed Y values* were complete. For example, suppose we are estimating a bivariate linear regression with *n* cases of which $n_1$ have observed values for *Y*. Then under MI, $v_{com,MI}$ would be $n - 2$, while under MID, $v_{com,MID}$ would be $n_1 - 2$.

Some readers may be misled into thinking that MI confidence intervals have more degrees of freedom than MID confidence intervals. This is not the case. Compared to MID, MI has more *complete*-data degrees of freedom $v_{com}$, but MI also has a larger fraction of missing information γ (see equation (34) below). Since the observed degrees of freedom is approximately $v_{obs} = (1-\gamma) \, v_{com}$, it follows that the observed degrees of freedom are typically *smaller* under MI than under MID.

## 4.3 Comparison

Asymptotically—that is, with an infinite number of imputations—MI and MID give equivalent point estimates, standard errors, and confidence intervals. When the number of imputations is limited, however, MID estimates are more precise than MI estimates.

Broadly speaking, the advantages of MID comes from the fact that MID estimates rely less on imputed values, and are therefore less affected by how much the imputed values vary from one data set to another. More precisely, MID reduces the between-imputation variance, which is the main source of uncertainty in MI estimates. To put the advantage of MID another way, MID reduces the fraction of missing information—the proportion of the total variance that lies between imputations—and the fraction of missing information determines the precision of point estimates, standard-error estimates, and confidence intervals.

### 4.3.1 Asymptotic comparison

We can better understand the difference between MI and MID if we spell out what quantities are being estimated. From a Bayesian perspective, imputation is a way to approximate a summary of the posterior density (Little and Rubin 2002). The posterior density is just the likelihood times the prior density $p(\theta)$ of $\theta$:

$$p(\theta \,|\, X_{obs}, Y_{obs}) = p(\theta)L(\theta \,|\, X_{obs}, Y_{obs}) \tag{22}.$$

If the prior is reasonably flat, then the posterior density is very similar to the likelihood.

Given this posterior density, the posterior mean of $\theta$ is

$$\overline{\theta} = E(\theta \mid X_{obs}, Y_{obs}) \tag{23},$$

and the posterior variance is

$$T = V(\theta \mid X_{obs}, Y_{obs}) \tag{24},$$

so that the posterior standard deviation is $\sqrt{T}$.

If the prior is reasonably flat and the likelihood is symmetric and unimodal (as we usually assume), then the posterior mean and standard deviation are for practical purposes indistinguishable from the maximum likelihood estimate and its standard error.

As the number of imputations $M$ gets large, MI and MID point estimates become equivalent since, in a Bayesian sense, both MI and MID point estimates are consistent estimators of the posterior mean:

$$\hat{\theta}_{M,MI} \xrightarrow[M \to \infty]{} \theta_{\infty,MI} = \theta_{\infty,MID} = \overline{\theta} = E(\theta \mid X_{obs}, Y_{obs}) \tag{25}.$$

Likewise, MI and MID standard-error estimates are asymptotically equivalent since both, when squared, are consistent estimators of the posterior variance:

$$\hat{T}_{M,MI} \xrightarrow[M \to \infty]{} T_{\infty,MI} = T_{\infty,MID} = T = V(\theta \mid X_{obs}, Y_{obs}) \tag{26}.$$

The *components* of the total variance, however, estimate different quantities under MID than they do under MI.

Under MID, the within- and between-imputation variances estimate the expectation of the posterior variance and the variance of the posterior expectation, conditionally on the observed $Y$ values, the observed $X$ values, and the missing $X$ values:

$$\hat{W}_{M,MID} \xrightarrow[M \to \infty]{} W_{\infty,MID} = E_{X_{mis}}[V(\theta \mid X_{obs}, Y_{obs}, X_{mis})] \tag{27}$$

$$\hat{B}_{M,MID} \xrightarrow[M \to \infty]{} B_{\infty,MID} = V_{X_{mis}}[E(\theta \mid X_{obs}, Y_{obs}, X_{mis})] \tag{28}.$$

Under MI, the within and between components estimate similar quantities, but the posterior mean and variance are conditioned on the missing values of $Y$ as well as $X$:

$$\hat{W}_{M,MI} \xrightarrow[M \to \infty]{} W_{\infty,MI} = E_{X_{mis}, Y_{mis}}[V(\theta \mid X_{obs}, Y_{obs}, X_{mis}, Y_{mis})] \tag{29}$$

$$\hat{B}_{M,MI} \xrightarrow[M \to \infty]{} B_{\infty,MI} = V_{X_{mis}, Y_{mis}}[E(\theta \mid X_{obs}, Y_{obs}, X_{mis}, Y_{mis})] \tag{30}.$$

Since conditioning on an extra variable can never increase the conditional variance (Wooldridge 2002), p. 31, it follows that

$$V(\theta \mid X_{obs}, Y_{obs}, X_{mis}, Y_{mis})] \leq V(\theta \mid X_{obs}, Y_{obs}, X_{mis}) \tag{31},$$

so that the within-imputation variance is smaller under MI than under MID:

$$W_{\infty,MI} \leq W_{\infty,MID} \tag{32}.$$

We foreshadowed this result earlier. The within-imputation variance is the mean square of the standard-error estimate from a single imputed data set. Deleting cases increases the standard error, so it makes sense that the within-imputation standard error would be smaller, on average, under MI than under MID.

Since the within-imputation variance is smaller under MI than under MID ($W_{\infty,MI} \leq W_{\infty,MID}$), and since the total variance is the same under both methods ($T_{\infty,MI} = T_{\infty,MID}$), it follows that the between-imputation variance must be larger under MI than under MID:

$$B_{\infty,MI} \geq B_{\infty,MID} \tag{33}.$$

Again, we foreshadowed this. The MI point estimate $\hat{\theta}_{MI}^{(m)}$ uses more random imputed values and therefore varies more across imputed data sets than does the MID point estimate $\hat{\theta}_{MID}^{(m)}$. So the between-imputation variance is larger under MI than under MID.

In sum, MI and MID estimates have the same asymptotic variance, but differ in how they split up the variance within and between imputed data sets. Under MID, more of the variance lies within the imputed data sets, and less lies between.

Since the fraction of missing information is the ratio of between variance to total variance, it it follows that the fraction of missing information is asymptotically smaller under MID than under MI:

$$\gamma_{\infty,MID} \leq \gamma_{\infty,MI} \tag{34}.$$

This inequality stands to reason, because the fractions $\gamma_{\infty,MI}$ and $\gamma_{\infty,MID}$ represent different quantities. Under MI, the fraction of missing information $\gamma_{\infty,MI}$ compares the available information to the information that would be present if *all* the cases were complete. Under MID, by contrast, the fraction of missing information $\gamma_{\infty,MID}$ compares the available information to the information that would be present if *only the cases with observed Y* were complete. Less information is missing by the latter standard than by the former, so the fraction of missing information is smaller under MID than under MI.

### 4.3.2 Efficiency of MI estimates with limited imputations

When the number of imputations is limited, imputation-based estimates are better if the fraction of missing information is small. Since the fraction of missing information is generally smaller under MID than under MI, it follows that MID estimates are more precise than MI estimates.

To see the advantages of having a small fraction of missing information, consider first the efficiency of point estimates. Under MI, the standard error of a point estimate based on *M* imputations is about

$$\frac{\gamma_{\infty,MI}}{2M} \times 100\% \tag{35}$$

percent larger than it would be with infinite imputations.[6] So with $\gamma_{\infty,MI} = 50\%$ missing information, $M = 5$ imputations are required to bring the standard error within 5% of its minimum possible value. But with $\gamma_{\infty,MI} = 20\%$ missing information, just $M = 2$ imputations will achieve the same goal.

---

[6] Expression (35) comes from von Hippel (2005). It approximates a more-complicated formula from Rubin (1987).

A similar result holds for estimating standard errors. Under MI, the true standard error of the finite-$M$ point estimate $\hat{\theta}_{M,MI}$ is $\sqrt{T_{\infty,MI}(1+\gamma_{\infty,MI}/M)}$ (Rubin 1987). But with a finite number of imputations $M$, this true standard error is unknown and has to be estimated by $\sqrt{\hat{T}_{M,MI}}$. When the fraction of missing information is large, $\sqrt{\hat{T}_{M,MI}}$ can be an unreliable estimate, in the sense that a noticeably different value of $\sqrt{\hat{T}_{M,MI}}$ would be obtained if the data were re-imputed. From one set of imputed data sets to another, the standard deviation of the squared standard-error estimate $\hat{T}_{M,MI}$ is about

$$\gamma_{\infty,MI}\sqrt{\frac{2}{M-1}} \tag{36}$$

times the square of the true standard error. (See Appendix A.) For example, with 50% missing information and $M = 5$ imputations, it would not be uncommon for the squared standard-error estimate $\hat{T}_{M,MI}$ to stray as much as 26% from the true value of the squared standard error. In other words, it would not be uncommon for the estimated standard error to stray as much as 12% from the true standard error.[7]

Again, MID can make the standard-error estimate more accurate by reducing the fraction of missing information.

---

[7] 1.12 is the square root of 1.26, so a 26% difference in the squared standard error corresponds to the 12% difference in the standard error.

MI confidence intervals can also be unreliable when the fraction of missing information is high (Royston 2004). Confidence intervals are affected not only by variation in the point estimate $\hat{\theta}_{M,MI}$ and in the standard-error estimate $\sqrt{\hat{T}_{M,MI}}$; they are also affected by variation in the degrees-of-freedom estimate $\hat{v}_{M,MI}$. When the fraction of missing information is large, Royston (2004) suggests that as many as $M = 20$ imputations may be required to produce confidence intervals that are reliable in the sense that they would not change substantially if the data were imputed again.

In addition to being unreliable, MI confidence intervals can have low coverage and excessive length when the fraction of missing information is large compared to the number of imputations. For example, with $\gamma_{\infty,MI} =50\%$ missing information and $M = 2$ imputations, a nominal 95% confidence interval will have just 92% coverage (Rubin and Schenker 1986). In addition, with $\gamma_{\infty,MI} =50\%$ missing information and $M = 2$ imputations, a confidence interval will have just $\hat{v}_{M,MI} =4$ degrees of freedom, which makes it 42% longer than it would be if the number of imputations were large.[8] If the number of imputations is increased to $M = 5$, the coverage would rise to 94.5%, but the interval would still be, on average, 8% longer than it would be if the number of imputations were infinite.

---

[8] As the degrees of freedom grows larger, the *t* distribution approaches the normal distribution, whose 97.5th percentile is 42% smaller than the corresponding percentile of a *t* distribution with 4 degrees of freedom.

### 4.3.3   Efficiency of MID estimates

Under MID, point estimates, standard errors, and confidence intervals depends on the MID fraction of missing information $\gamma_{\infty,MID}$ in the same way that MI estimates depend on the MI fraction of missing information $\gamma_{\infty,MI}$. Since the fraction of missing information is generally smaller under MID than under MI, MID estimates are generally more efficient than MI estimates. The advantages of MID may be expressed in either of the following ways:

1. With the same number of imputations, estimates will generally be more accurate under MID than under MI.

2. Or: to achieve a desired level of precision, fewer imputations are needed under MID than under MI.

Reducing the number of required imputations can be a substantial benefit when each imputed data set requires considerable time and resources. For example, in Downey et al.'s (2004) complicated analysis of a large data set, it took about an hour to impute and analyze a single imputed data set. MID reduced the number of imputations needed from ten to three, which saved weeks of research time over the course of the study.

## 5.  HOW MUCH DOES MID HELP?

## A SIMULATION EXPERIMENT

The analytic results in the previous section show that MID estimates are more efficient than MI estimates, but they do not completely show whether the benefits of MID are large in practice. In this section, we make our findings more concrete through a simulation experiment that compares the properties of MI and MID estimates under a variety of circumstances. We then extend the experiment to explore the implications of deleting cases before rather than after imputation.

## 5.1   Design

In each simulated data set, we generated $N$ observations on two input variables $(X_1, X_2)$; these variables were bivariate standard normal with a correlation of $\rho_{12}$. We then generated the outcome variable $Y$ from the following regression model:

$$Y = \alpha + \beta_1 X_1 + \beta_2 X_2 + e,$$

$$\text{where } e \sim \text{Normal } (0, \sigma_e^2) \tag{37}.$$

We deleted values of $X_2$ and $Y$ in one of three ignorable patterns described below. After deletion, we created $M$ copies of the incomplete data set, and we imputed missing values under a multivariate normal model. To impute missing values, we used the MI procedure in SAS 9.1. The procedure options were set so that the program used up to 1000 iterations of the EM method to find the posterior mode, then used Markov-chain Monte Carlo (MCMC) to generate the actual imputations.

In carrying out the simulations, we independently varied six different factors:

1. $N$, the number of observations in each data set, took values of 50 and 200.

2.  $\rho_{12}$, the correlation between $X_1$ and $X_2$, took values of .2, .5 and .8.

3.  $R^2$, the proportion of variance explained in $Y$, also took values of .2, .5 and .8. (We manipulated $R^2$ by leaving $(\alpha, \beta_1, \beta_2)$ alone and setting

    $\sigma_e^2 = 2 (1 - R^2) (1 + \rho_{12}) / R^2$.)

4.  $p$, the proportion of $X_2$ and $Y$ values that we deleted, took values of .2 and .5.

5.  We deleted values in one of three ignorable patterns:

    a.  *Missing completely at random (MCAR).*

        $X_2$ was deleted with constant probability $p$.

        $Y$ was independently deleted with the same probability.

    b.  *Coordinated missingness.*

        $X_2$ and $Y$ tend to be missing from the same cases.

        Specifically, $X_2$ was deleted with probability $2 p \Phi(X_1)$,

        and $Y$ was independently deleted with the same probability

        Here $\Phi$ is the cumulative standard normal density.

    c.  *Complementary missingness.*

        $X_2$ and $Y$ tend to be missing from different cases.

        $X_2$ was deleted with probability $2 p \Phi(X_1)$, and

        $Y$ was deleted with probability $2 p \Phi(-X_1)$.

6.  Finally, $M$, the number of imputations, took values of 2, 5, and 10.

The intercept and slopes of the regression model were fixed at $(\alpha, \beta_1, \beta_2) = (1,1,1)$. It was not necessary to vary the intercept and slopes since changing $(\alpha, \beta_1, \beta_2)$ to any nonzero value would be equivalent to changing $\sigma_e^2$ and shifting or rescaling the axes. And the experiment already manipulated $\sigma_e^2$ as a way of manipulating $R^2$.

For each of the 324 combination of the six experimental factors, we simulated $D$=1000 data sets. Then, using an ordinary MI strategy, we obtained point estimates and nominal 95% confidence intervals from the imputed data. Next we deleted cases with imputed $Y$ values, and obtained MID estimates from the cases that remained.

Note that we did not re-impute the data before using MID. Instead, to make the results as comparable as possible, we based our MI and MID estimates on the same imputed data sets.

## 5.2   Results

We compared MI to MID using the length and coverage of confidence intervals and the absolute estimation error of point estimates. Coverage was defined conventionally. The other comparisons were defined as follows:

- *Length of confidence intervals.* For a given parameter $\theta$ and data set $d$, let $\lambda_{d,MI}$ and $\lambda_{d,MID}$ be the lengths of nominal 95% confidence intervals obtained by MI and MID. Then the percent difference in length is $100\% \times \left( \lambda_{d,MI} - \lambda_{d,MID} \right) / \lambda_{d,MI}$.

- *Absolute error of point estimates.* Likewise, let $\hat{\theta}_{d,MI}$ and $\hat{\theta}_{d,MID}$ be MI and MID

point estimates of the parameter $\theta$ in data set $d$. Letting the absolute errors of the estimates be $|\varepsilon_{d,MI}| = |\hat{\theta}_{d,MI} - \theta|$ and $|\varepsilon_{d,MID}| = |\hat{\theta}_{d,MID} - \theta|$, the percent difference in absolute error is $100\% \times (|\varepsilon_{d,MID}| - |\varepsilon_{d,MI}|)/|\varepsilon_{d,MI}|$.

Note that these comparisons account for the pairing of MI and MID estimates that come from the same data set $d$. Note also that the percentage comparisons are inherently asymmetric; the minimum percent difference is –100%, but there is no maximum percent difference since the denominator of the comparison can be close to zero. Because of this asymmetry, we summarize the comparisons by using the median instead of the mean.

Across all the simulated data sets, MID had a small advantage over MI. Nominal 95% confidence intervals had a coverage rate of 92.8% under MI and 93.8% under MID; that is, the coverage of MID confidence intervals was, on average, 1% higher and 1% closer to the nominal rate. The median difference in the length of confidence intervals was –2.75%, and the median difference in absolute estimation error was –2.4%. That is, in half of all data sets, the MID confidence interval was at least 2.75% shorter than the MI confidence interval, and the MID estimate was at least 2.4% closer to the true parameter value than was the MI estimate.

Table 1 summarizes the experimental results. For concision's sake, Table 1 collapses the results across experimental factors that had little effect on the comparison between MI and MID. As it turns out, only two factors made a material difference; these factors were the proportion of missing values, and the number of imputations. Table 1 gives results for all four regression parameters ($\alpha$, $\beta_1$, $\beta_2$, $\sigma^2$); in addition, Figure 1 summarizes the results for the first slope $\beta_1$.

**←TABLE 1 NEAR HERE→**

**←FIGURE 1 NEAR HERE→**

The advantages of MID were greatest when there were few imputations and a lot of missing values; in addition, MID seemed to have a greater advantage for estimating the intercept $\alpha$ than for estimating the other regression parameters. When estimating the intercept under the most difficult experimental settings ($M$=2, $p$=.5), MID confidence intervals had 5% better coverage and were, at the median, 26% shorter than MI confidence intervals; in addition, MID point estimates were, at the median, 9% closer to the parameter value than were MI point estimates. In percentage terms, MID did more to shorten confidence intervals than it did to increase coverage or to improve point estimates. This may be because the length of confidence intervals has three sources of imputation error: error in the point estimate, error in the standard error estimate, and error in estimating the $t$ statistic's degrees of freedom. All three errors are reduced by MID.

When there were more imputations or fewer missing values, the differences between MI and MID were relatively modest. But under all experimental settings, MID was at least as good as MI. Although there was one setting ($M$=2, $p$=.2) where MI produced 1–2% shorter confidence intervals than MID, under that setting MI's advantage in length came at the cost of a 1% disadvantage in coverage.

## 5.3   Deletion *before* imputation?

In the introduction, we said that deletion should come *after* imputation, because cases with missing $Y$ may be useful for imputing $X$ in other cases. But how useful, in practice, are cases with missing $Y$? Little (1992, p. 1227), suggested that cases with missing $Y$ contain just a "minor amount of information for the regression of interest." Does it really matter whether we delete cases before or after imputation?

To find out, we tested a strategy of *deletion, then multiple imputation* (DMI), under which we deleted cases with missing $Y$, then used the remaining cases for imputation and analysis. We included this method in the six-factor experiment described above.

Under most experimental conditions, DMI and MID produced very similar results. Across all the simulated data sets, DMI point estimates were, at the median, just 0.7% further from the true parameter values than were MID point estimates. Similarly, DMI confidence intervals had just 0.1% lower coverage and were, at the median, just 0.4% longer than MID confidence intervals,

Under selected conditions, however, DMI had more of a disadvantage. Specifically when there were 50% missing values, and the pattern of missingness was *complementary*—that is, values of $X_2$ were missing from different cases than values of $Y$—DMI became a seriously flawed strategy. Under these circumstances, deleting cases with $Y$ missing meant deleting most of the cases with $X_2$ observed. With few observed $X_2$ values, it became very difficult to impute $X_2$ precisely, so that on occasion (about once in a thousand data sets), it was impossible to impute values at all. When values could be obtained, the DMI point estimates were a median of 4–5% further from the true parameter values than were the MID point estimates. Likewise DMI confidence intervals were a median of 3–4% longer than MID confidence intervals, with 0.1–0.7% lower coverage. These differences were about the same whether the number of imputations was $M$=2, 5, or 10.

In short, under most circumstances, DMI is nearly as good as MID, but the advantages of MID can be substantial when the cases missing $X$ overlap little with the cases that are missing $Y$.

# 6. EXTENSIONS

For simplicity's sake, we have focused on estimating individual parameters for the conditional distribution of a single $Y$ variable. The results, however, can easily be extended to the situation where $Y$ is multivariate or when inference concerns multiple parameters.

## 6.1 Extension to multiple parameters

Researchers sometimes wish to carry out multi-parameter hypothesis tests—for example, a test of the null hypothesis $H_0$: $(\alpha, \beta_1, \beta_2)$=(0,0,0). Three methods have been proposed for multi-parameter inference with multiply-imputed data. One method (Li et al. 1991) is a generalization of the single-parameter inference in section 4, which is justified in terms of the posterior density. The other methods are based on likelihood theory (Li et al. 1991; Meng and Rubin 1992).

Since MID estimates the same likelihood and posterior as MI (see section 3), we would expect that the established multi-parameter tests could be used under MID with little or no modification and some advantage in efficiency.

## 6.2   Extension to multivariate $Y$: for example, repeated measures

The discussion so far has focused on the situation where $Y$ is univariate, but it is not hard to see the implications when the dependent variable is a multivariate vector $Y = (Y_1, Y_2, \ldots)$. Under MID, the basic prescription is to delete imputed elements of $Y$ before analysis, unless this would require deleting observed elements of $Y$ as well.

An important application occurs when $Y$ contains repeated measures. For example, in the Early Childhood Longitudinal Study, Kindergarten cohort, the vector $y_i = (y_{i1}, y_{i2}, y_{i3}, y_{i4})$ represents student $i$'s scores on four reading tests, taken near the beginning and end of kindergarten and first grade. By design, 70% of children are missing the third test score $y_{i3}$, and many children are missing other test scores as well.

In the imputation step, missing parts of the $y_i$ vector can be filled in along with missing values of other variables $x_i$. With respect to imputed $y_i$ values, this means that there are three types of student:

(1) For some students, the entire $y_i$ vector is observed.

(2) For some students, the entire $y_i$ vector is imputed.

(3) For the remaining students, only part of the $y_i$ vector—e.g., $y_{i3}$—is imputed and the remaining parts, e.g., $(y_{i1}, y_{i2}, y_{i4})$—are observed.

Under MID, it is clear that students in the first group would be retained after imputation, while students in the second group would be deleted. The treatment of the third group, however, depends on the analysis method that used by the researcher.

(1) Some older methods—e.g., repeated-measures MANOVA (Potthoff and Roy 1964)—require "balanced" data with values for every component of the $y_i$ vector. If such methods are used, then it may be advisable to use partly imputed values for $y_i$. For example, the analysis could use an imputed value for $y_{i3}$ along with the observed values of ($y_{i2}$, $y_{i2}$, $y_{i4}$). Using imputed $y_{i3}$ values involves a tradeoff, however, since with a finite number of imputations, variation in the imputed $y_{i3}$ value may offset the efficiency gained by using the observed values of ($y_{i1}$, $y_{i2}$, $y_{i4}$).

(2) Newer methods—e.g., multilevel growth models (Raudenbush 2001)—can handle "unbalanced" data where some students are missing parts of $y_i$. If such methods are used, then there is no need to include imputed elements of $y_i$ in the analysis. For example, in a case with $y_{i3}$ missing, an MID analysis would use the observed values of ($y_{i1}$, $y_{i2}$, $y_{i4}$) but would not require an imputed value for $y_{i3}$.

Downey et al. (2004) used the second approach to estimate learning rates in the Early Childhood Longitudinal Study, Kindergarten cohort.

## 7. CAN MID BE WORSE THAN A CONVENTIONAL MI STRATEGY?

The justification for MID depends on the claim that imputed $Y$ values contain no information about the regression of interest. Although this claim is often correct, there are situations where the imputed $Y$ values do contain some extra information. In these situations, a conventional MI strategy can potentially be superior to MID. The extra information in the imputed values must be substantial, however, or the benefits of the extra information will be swamped by random variation in the imputed values.

## 7.1  Auxiliary variables

One situation where imputed $Y$ values can contain extra information is when *auxiliary variables* have been used in imputation. An auxiliary variable is one that, though not part of the intended analysis, can improve imputation by providing extra information about the incomplete variables. For example, the intended analysis may be a regression of $Y$ on $X_1$ and $X_2$, but the imputation step also uses an auxiliary variable $Z$ to improve imputation of $Y$.

Note that auxiliary variables are irrelevant to the MI-MID comparison unless the auxiliary variables improve the imputation of $Y$. If only the imputed $X$ values benefit from auxiliary information, then MI and MID estimates will benefit equally, and MID will continue to have an advantage over MI. It is only when the imputed $Y$ values benefit from auxiliary information that MI can outperform MID.

When auxiliary variables improve imputation of $Y$, we might expect MI to be more efficient MID—and, asymptotically, it will be. With an infinite number of imputations, MI with auxiliary variables will produce "superefficient" estimates whose asymptotic standard errors are smaller than the asymptotic standard errors $\sqrt{T_{\infty,MI}}$ obtained from a conventional MI analysis (Meng 1995; Rubin 1996). Since MID's asymptotic standard errors $\sqrt{T_{\infty,MID}}$ are equal to those obtained from a conventional MI strategy—i.e., since $\sqrt{T_{\infty,MI}} = \sqrt{T_{\infty,MID}}$ —we would expect that, with an infinite number of imputations, MI with auxiliary variables will have a smaller standard error than MID.

But in practice, we cannot use an infinite number of imputations—and with a limited number of imputations, even auxiliary variables may not eliminate MID's advantage over MI. With a limited number of imputations $M$, MI's smaller infinite-$M$ variation can be overwhelmed by the random variation that comes from using a finite number of imputed values for $Y$.

To compare the finite-$M$ properties of MI and MID in the presence of auxiliary variables, we extended our simulation experiment. In the modified experiment, we included an auxiliary variable $Z$ which had a correlation $\rho_{YZ}$ with $Y$ but no further relationship with $X$. Specifically, we generated $Z = \rho_{YZ} Y_s + u$, where $Y_s$ is a standardized version of $Y$, and $u \sim N(0, 1 - \rho_{YZ}^2)$ is a normal disturbance that is independent of $X_1$, $X_2$, and $Y$. The correlation $\rho_{YZ}$ took values of $\{.1, .3, .5, .7, .9\}$ and was varied independently of the other six factors in the simulation experiment. As in the original experiment, we simulated $D=1000$ data sets for each combination of factors, then compared MI and MID with respect to the length and coverage of confidence intervals and the absolute error of point estimates.

<div align="center">←<b>FIGURE 2 NEAR HERE</b>→</div>

Figure 2 summarizes the results for the slope $\beta_1$; results for other parameters were similar. The results are plotted as a function of the correlation $\rho_{YZ}$, the fraction of missing information $p$, and the number of imputations $M$. As before, we collapsed the results across the other four experimental factors, which had comparatively small effects.

When an auxiliary variable $Z$ was used, the differences between MI and MID depended on the circumstances.

- In some circumstances, the advantages of MID were substantial. Under the extremest settings—with 50% missing values, just a $\rho_{YZ}$=.1 correlation between $Y$ and $Z$, and only 2 imputations—the MID confidence intervals had 4.6% better coverage (92.6% vs. 88.0%) and were a median of 7.1% shorter than MI confidence intervals. In addition, MID point estimates were a median of 22% closer to the parameter values than were MI point estimates.

- But at the opposite extreme—with 50% missing values, 10 imputations, and a $\rho_{YZ}$=.9 correlation between $Y$ and $Z$—the advantage tipped heavily toward MI. Under these circumstances, MI point estimates were a median of 22% closer to the parameter value than MID point estimates. Similarly, MI confidence intervals were a median of 12.5% shorter than MID confidence intervals. Although coverage was slightly lower under MI than under MID (95.8% vs. 94.3%), both methods were about equally close to the nominal coverage rate of 95%.

With 20% missing values, the patterns were similar, but the differences between MI and MID were much smaller.

In sum, the auxiliary information has to be quite good before it trumps the extra variation introduced by using a finite number of randomly imputed $Y$ values. In the experiment, the tipping point where MI became more efficient than MID was around $\rho_{YZ}=.5$ with $M=10$ imputations, around $\rho_{YZ}=.6$ with $M=5$ imputations, and around $\rho_{YZ}=.7$ with $M=2$ imputations.[9] In addition, the simulation may favor the auxiliary-variable approach in the sense that the simulated auxiliary variable $Z$ has no missing values. In a real data set, the same cases that were missing $Y$ would often be missing $Z$ as well—so that, even if the correlation $\rho_{YZ}$ was large, $Z$ could do little to improve imputation of $Y$. As a practical matter, then, MI with auxiliary variables may be a poorer strategy than Figure 2 suggests.

## 7.2   Nonignorable missingness

Another situation where the imputed $Y$ values contain extra information is when $Y$ has been imputed under a *nonignorable* model. In previous sections, we assumed that the mechanism which causes $Y$ to be missing is *ignorable* in the sense that, in cases where $Y$ is missing, the unobserved value of $Y$ is similar to the observed $Y$ values in other cases with similar values for $X$ (Little and Rubin 2002). But this assumption is not always credible. For example, if $Y$ is body weight, we might suspect that nonrespondents have higher $Y$ values than do respondents with comparable values for $X$ variables such as gender, height, and age.

A simple way to adjust imputations for nonignorability is to impute $Y$ under an ignorable model, and then adjust the imputed values to compensate for the presumed nonresponse bias (Rubin 1987). For example, if $Y$ is self-reported body weight, we might adjust the imputed $Y$ values upward by, say, 10 pounds. Then the imputed $Y$ values contain information that is not available from other cases, and this information would be lost if cases with imputed $Y$ were deleted before analysis. MID is inappropriate in this situation. Similarly, MID is inappropriate if imputed $X$ values have been adjusted in cases that also have imputed $Y$s.

---

[9] With $M=2$ imputations, however, the coverage of MI confidence intervals was almost 5% below the nominal rate, so that the MID confidence intervals may be preferable even though they are longer.

Again, however, the extra information in the imputed $Y$ values must be quite good before it can improve the estimates. In many practical settings, the form of nonignorable missingness is not understood well enough to be useful for improving the imputations. When using MI, few analysts adjust imputations for nonignorability, and when adjustments are made, the adjustments can make estimates worse instead of better (Rubin 2003).

In short, although MID lacks the tools to adjust for certain types of nonignorability, those tools can be dull, and in practice they are usually left in the shed.

## 7.3   MID is limited to estimating conditional distributions

Finally, we should emphasize that that MID is only valid for estimating the parameters $\boldsymbol{\theta}_{Y|X}$ that govern the conditional distribution of $Y$ given $X$. MID will not necessarily produce good estimates for the parameters $\boldsymbol{\theta}_X$ that govern the distribution of the $X$ variables, or for the parameters $\boldsymbol{\theta}_Y$ that govern the marginal distribution of $Y$. For example, if we are estimating the regression of $Y$ on $X$, MID will produce consistent estimates for the intercept $\alpha$, slopes $\boldsymbol{\beta}$, and residual variance $\sigma_e^2$—but MID may not produce a good estimate of $R^2$. This is because, while $\alpha$, $\beta$, and $\sigma_e^2$ describe the conditional distribution of $Y$, calculating $R^2$ also requires information about the marginal distribution of $X$ or $Y$. For example, one $R^2$ formula—

$R^2 = 1 - \sigma_e^2/\sigma_Y^2$—requires the marginal variance $\sigma_Y^2$ of $Y$. A valid estimate for $R^2$ can be

obtained by combining an MI estimate of $\sigma_Y^2$ with an MID estimate of $\sigma_e^2$, or (less accurately but more simply) by applying an ordinary MI strategy.

## 8. CONCLUSION

We have proposed a modified strategy for obtaining regression estimates when both $X$ and $Y$

are missing values at random. The strategy is called *multiple imputation, then deletion*

(*MID*), and consists of imputing the data as usual, but then deleting cases with imputed $Y$

values before analysis.

When there is something wrong with the imputed $Y$ values, MID offers protection from the

consequences of using the problematic values in analysis. And when the imputed $Y$ values

are acceptable, MID still typically offers somewhat greater efficiency than an ordinary MI

strategy.

The justification for MID is that, under typical circumstances, imputed $Y$ values contain no information about the regression of interest, so that imputed $Y$ values add nothing but noise to the estimates. There are special circumstances where the imputed $Y$ values do contain some extra information, but that information must be quite substantial before it compensates for the noise that results from using imputed $Y$ values in analysis.

MID is especially attractive when there are a lot of missing $Y$ values and it is difficult to specify a convincing imputation model for $Y$. Since MID is a simple variant of MI, MID is not hard to implement, as illustrated by the short SAS macro in Appendix B.

# APPENDIX A:

# RELIABILITY OF THE STANDARD-ERROR ESTIMATE

Expression (36) gives the reliability of the squared standard-error estimate $\hat{T}_{M,MI}$ under MI. The derivation of that expression is given here.

Over all sets of $M$ imputations, the expectation and variance of $\hat{T}_{M,MI}$ are

$$E(\hat{T}_{M,MI}) = (1 + \gamma_{\infty,MI} / M)T_{\infty,MI} \tag{38}$$

and

$$\begin{aligned} V(\hat{T}_{M,MI}) &= V(\hat{W}_{M,MI}) + V(\hat{B}_{M,MI}) \\ &\approx V(\hat{B}_{M,MI}) = 2B_{\infty,MI}^2 /(M-1) \end{aligned} \tag{39}$$

(Rubin 1987). The term $V(\hat{W}_{M,MI})$ can be neglected since $V(\hat{W}_{M,MI}) << V(\hat{B}_{M,MI})$ if $n$ is large.[10]

From (38) and (39), we can derive an expression for $\hat{T}_{M,MI}$'s coefficient of variation,

---

[10] Again, Rubin's (1987) expression for $\hat{B}_{M,MI}$ omits the correction factor $(1+1/M)$. To compensate, he includes this correction in his expression for the variance of $\hat{T}_{M,MI}$:

$$V(\hat{T}_{M,MI}) = 2(1+1/M)^2 B_{\infty,MI}^2 /(M-1).$$

$$\frac{\sqrt{V(\hat{T}_{M,MI})}}{E(\hat{T}_{M,MI})} = \gamma_{\infty,MI}\left(\frac{M+1}{M+\gamma_{\infty,MI}}\right)\sqrt{\frac{2}{M-1}} \approx \gamma_{\infty,MI}\sqrt{\frac{2}{M-1}} \qquad (40),$$

which equals expression (36).

# APPENDIX B:

# SAS MACRO IMPLEMENTATION

Because MID is a very simple extension of MI, implementing MID is straightforward provided there already exists an implementation of MI. SAS, for example, has implemented the MI procedure for multiple imputation under an assumption of conditional normality.[11] The following SAS macro implements MID by

1. creating a binary variable (*y_missing*) that indicates which cases are missing *Y*;

2. calling the MI procedure for multiple imputation;

3. and deleting from the imputed data all cases where *y_missing* indicates that *Y* was missing before imputation.

The macro arguments are the name of the *incomplete_data* set, the desired name for the *imputed_data* set, the number of desired imputations (*nimpute*, default 5), the outcome variable (*y*), and the list of input variables (*xs*).

**Example of usage**

```
%mid (incomplete_data=your_incomplete_data,
   imputed_data=your_imputed_data,
   y=income, xs=education age,
   nimpute=10);
```

---

[11] The MI procedure can also impute binary variables if the pattern of missing values is monotone.

**Macro code**

```
%macro mid (incomplete_data=, imputed_data=, y=, xs=, nimpute=5);
 DATA &incomplete_data;
  SET &incomplete_data;
  if (&y=.) then y_missing=1; else y_missing=0;
 RUN;
 PROC MI DATA=&incomplete_data OUT=&imputed_data NIMPUTE=&nimpute;
  VAR &xs &y;
 RUN;
 DATA &imputed_data
  SET &imputed_data;
  if y_missing then delete;
 RUN;
%mend mid;
```

# TABLES

*Table 1*. Differences between MI and MID estimates

| Percent missing | Number of imputations | Parameter | Coverage of nominal 95% confidence intervals | | | Median difference in… | |
|---|---|---|---|---|---|---|---|
| | | | MI | MID | Difference | …length of confidence interval | …absolute error of point estimate |
| 50% | 2 | Intercept $\alpha$ | 89% | 94% | 5% | −26% | −9% |
| | | Slope $\beta_1$ | 89% | 93% | 4% | −25% | −7% |
| | | Slope $\beta_2$ | 88% | 91% | 3% | −18% | −4% |
| | | Residual variance $\sigma^2$ | 87% | 90% | 3% | −11% | −6% |
| | 5 | Intercept $\alpha$ | 94.2% | 94.7% | 0.5% | −13% | −3% |
| | | Slope $\beta_1$ | 94.1% | 94.5% | 0.4% | −12% | −3% |
| | | Slope $\beta_2$ | 93.1% | 93.4% | 0.3% | −9% | −2% |
| | | Residual variance $\sigma^2$ | 89.8% | 90.7% | 0.9% | −6% | −3% |
| | 10 | Intercept $\alpha$ | 94.8% | 94.8% | 0.0% | −7% | −2% |
| | | Slope $\beta_1$ | 94.7% | 94.8% | 0.1% | −6% | −2% |
| | | Slope $\beta_2$ | 93.8% | 93.9% | 0.2% | −4% | −1% |
| | | Residual variance $\sigma^2$ | 90.5% | 91.0% | 0.4% | −3% | −2% |
| 20% | 2 | Intercept $\alpha$ | 94.0% | 94.9% | 0.9% | +1% | −4% |
| | | Slope $\beta_1$ | 93.9% | 94.9% | 1.0% | +1% | −4% |
| | | Slope $\beta_2$ | 93.4% | 94.4% | 1.0% | +1% | −4% |
| | | Residual variance $\sigma^2$ | 92.6% | 93.4% | 0.9% | +2% | −4% |
| | 5 | Intercept $\alpha$ | 95.0% | 95.0% | 0.0% | −2% | −2% |
| | | Slope $\beta_1$ | 94.9% | 95.0% | 0.0% | −1% | −2% |
| | | Slope $\beta_2$ | 94.7% | 95.0% | 0.3% | −2% | −2% |
| | | Residual variance $\sigma^2$ | 93.2% | 93.4% | 0.2% | −1% | −2% |
| | 10 | Intercept $\alpha$ | 95.0% | 95.0% | 0.0% | −1% | −1% |
| | | Slope $\beta_1$ | 94.9% | 95.0% | 0.1% | −1% | −1% |
| | | Slope $\beta_2$ | 94.9% | 94.8% | −0.1% | −1% | −1% |
| | | Residual variance $\sigma^2$ | 93.5% | 93.6% | 0.1% | 0% | −1% |

*Caption*. Compared to MI estimates, MID point estimates tend to be closer to the parameter values, and MID confidence intervals tend to be shorter with equal or higher coverage. The advantages of MID are larger when there are few imputations or a lot of missing values.

# FIGURES

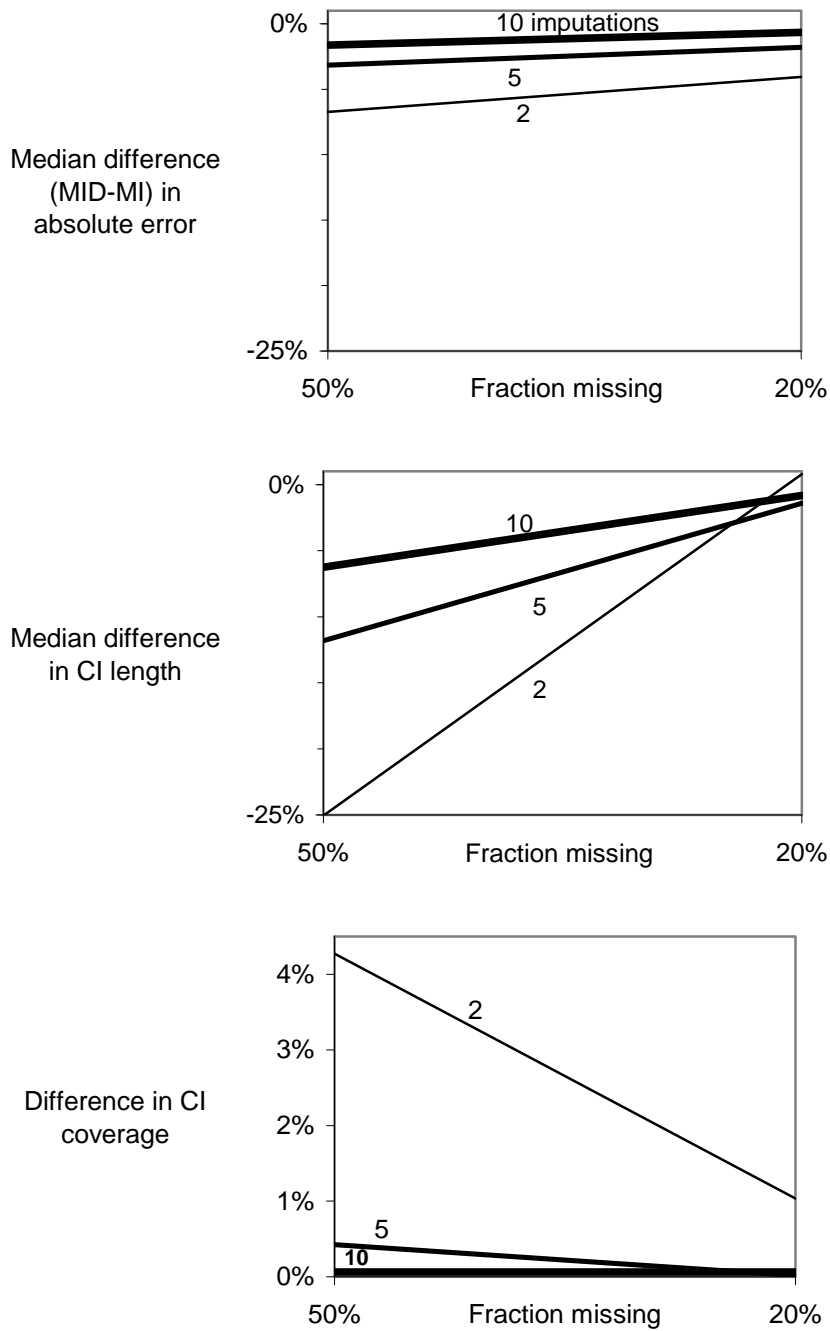

*Figure 1*. Compared to an ordinary MI strategy, MID typically produces smaller errors in point estimates, as well as shorter confidence intervals with higher coverage rates. These plots focus on estimation of the slope $\beta_1$; estimates for other parameters follow a similar pattern.

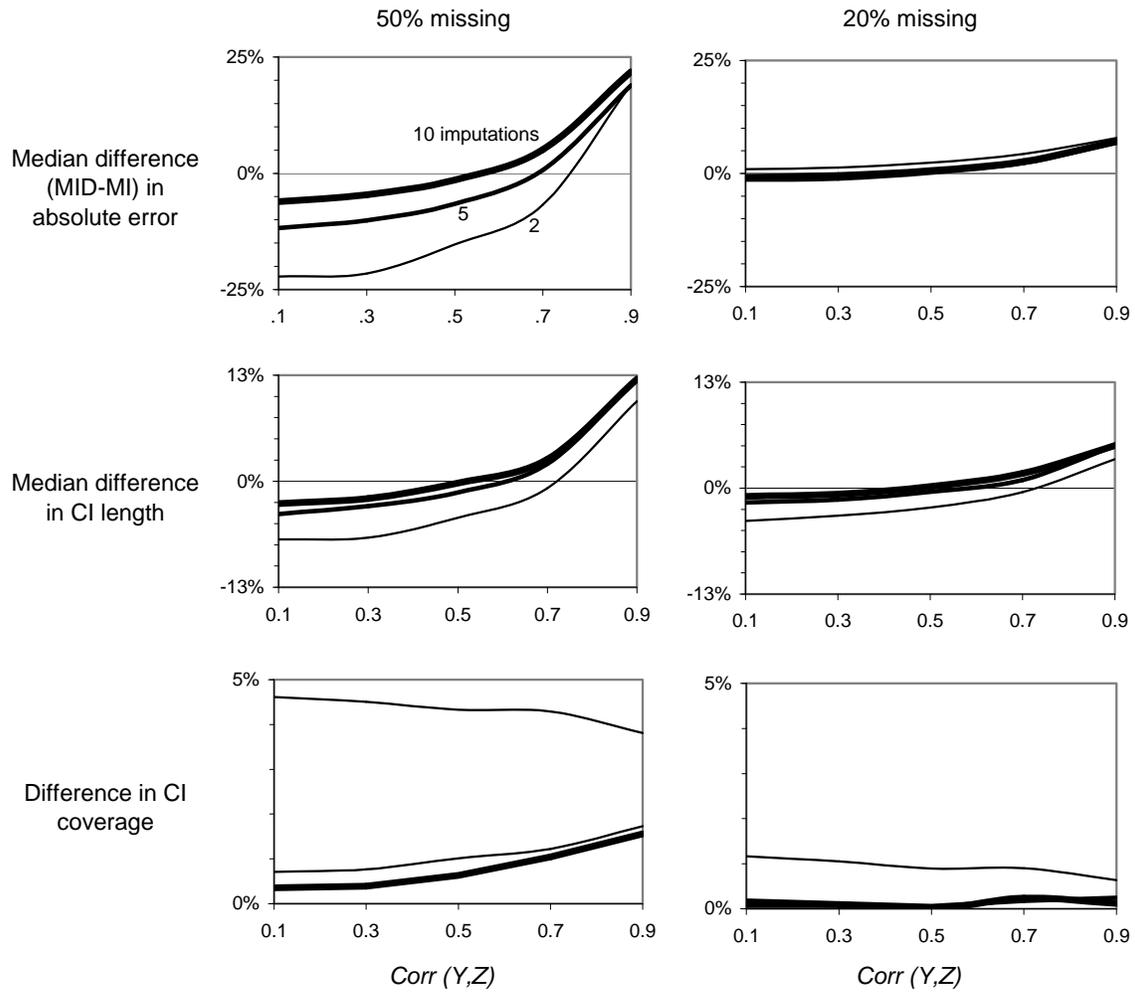

*Figure 2*. Differences between MID and MI when the imputations use an auxiliary variable *Z* that is correlated with *Y*. Unless the correlation is strong and the auxiliary variable is complete in cases where the *Y* variable is missing, the improved MI estimates are no better than the MID estimates. These plots focus on estimation of the slope $\beta_1$; estimates for other parameters follow a similar pattern.